\documentclass[%
 aip,
 amsmath,amssymb,
 preprint,%
]{revtex4-1}

\usepackage{graphicx}% Include figure files
\usepackage{dcolumn}% Align table columns on decimal point
\usepackage{bm}% bold math
\usepackage[utf8]{inputenc}
\usepackage[T1]{fontenc}
\usepackage{mathptmx}
\usepackage{etoolbox}

\usepackage{color}

\makeatletter
\def\@email#1#2{%
 \endgroup
 \patchcmd{\titleblock@produce}
  {\frontmatter@RRAPformat}
  {\frontmatter@RRAPformat{\produce@RRAP{*#1\href{mailto:#2}{#2}}}\frontmatter@RRAPformat}
  {}{}
}%
\makeatother
\begin{document}

\preprint{}

\title[Selective modification of inviscid invariants]{Stabilizing two-dimensional turbulent Kolmogorov flow via selective modification of inviscid invariants}
% Force line breaks with \\
\author{Gaurav Kumar}
\author{Aditya Nair}%
 \email{adityan@unr.edu.}
\affiliation{ 
Department of Mechanical Engineering, University of Nevada, Reno, Nevada, 89557%\\This line break forced with \textbackslash\textbackslash
}%

\date{\today}

\begin{abstract}
We stabilize two-dimensional turbulent Kolmogorov flow by selectively altering the time rate of change of inviscid invariants (energy and enstrophy) of the flow. This method has earlier been demonstrated to modify the two-dimensional unforced decaying turbulent flow to reach steady states. However, Kolmogorov flow exhibits flow unsteadiness over a range of spatial and temporal scales which are driven by constant external forcing at wave number $k_f$. The energy injected at scale ($k \approx k_f$) is distributed to large ($k > k_f$) and small ($k < k_f$) wave numbers through interactions of different scales called direct and inverse energy cascading, respectively. The selective modification strategy autonomously identifies additional forcing inputs into the governing equations which results in the transformation of Kolmogorov flow into a non-trivial steady state. We demonstrate that this method is highly effective in altering the inverse energy cascading, a particularly intricate challenge in fluid dynamics. Intriguingly, we have also found that the steady states achieved through this approach resemble an invariant solution of the original Kolmogorov flow, highlighting the potential significance of our method.
\end{abstract}

\maketitle

\section{Introduction}
 \label{sec:intro}

Turbulence is a phenomenon typically characterized by chaotic and irregular fluid motion, often observed in natural processes like wind, ocean currents, and atmospheric flows. One of the key features of turbulence is the energy cascade, which involves the transfer of energy across different scales of motion. In a turbulent flow, energy is continuously transferred from larger scales (smaller wave numbers) to smaller scales (larger wave numbers) through a process known as the direct energy cascade. In this process, larger eddies repeatedly break down into smaller eddies until the energy is eventually dissipated into heat \citep{pope2000turbulent}. However, in certain turbulent flows, an inverse energy cascade can also appear, where energy is transferred from smaller scales to larger scales \citep{paret1997experimental, biferale2012inverse, francois2013inverse}. This phenomenon is observed in geophysical flows and is responsible for the formation of large-scale structures such as hurricanes and atmospheric patterns. Hence, there is a huge interest in the research community to understand the energy transfer due to interaction among scales of motion in turbulence which is crucial for effective control of the turbulent flow for practical applications.

Kolmogorov flow, introduced by Kolmogorov in 1960 \citep{kolmogorov1960}, serves as a valuable tool for investigating turbulence and its transition in fluid flows. This flow is governed by the two-dimensional Navier-Stokes equation in a bi-periodic domain, featuring a sinusoidal body force in the cross-stream direction. Two key non-dimensional parameters, the Reynolds number ($Re$) and the forcing wave number ($n$), play a crucial role in determining the stability and transition characteristics of this flow. For large values of $n$ and $Re$, the solution to this flow is generally unstable.

In two-dimensional turbulent flow, kinetic energy and enstrophy (the mean squared vorticity) are the inviscid invariants. In cases of forced two-dimensional turbulent flow with a constant forcing rate at wave number $k_f = 2\pi n$, previous research has shown that energy tends to cascade towards smaller wave numbers ($k < k_f$), while enstrophy cascades towards larger wave numbers ($k > k_f$) \citep{kraichnan1967inertial, gupta2019energy}. Additionally, the work by \citet{shu1993effective} has highlighted the presence of an inverse energy cascade process in Kolmogorov flow as the forcing wave number $n$ approaches infinity. This study demonstrates that energy transfers from high-forcing wave numbers to lower wave numbers, resulting in the emergence of large-scale spatial-temporal coherent structures. Given these insights, modifying the flow of kinetic energy or enstrophy across scales within the Kolmogorov flow presents a promising avenue for effectively stabilizing this flow.

In a control-based approach by \citet{lucas2022stabilization}, a time-delayed feedback control combined with the knowledge of system symmetries was exploited to find exact coherent structures of Kolmogorov flow. In previous studies \cite{willis2017surfing, linkmann2020linear}, researchers stabilized invariant solutions of pipe and channel flows by employing linear feedback control to counteract edge instabilities. A notable challenge in implementing such control mechanisms, aimed at reducing flow unsteadiness selectively, lies in avoiding unintended laminarization of the flow. This often necessitates hyperparameter tuning in the context of previous control-based approaches, especially when the goal is to maintain a non-trivial steady state of the flow, as emphasized in \citet{willis2017surfing}.

In this study, our aim is to stabilize the Kolmogorov flow by modifying the underlying energy cascade mechanism. We employ the selective energy and enstrophy modification approach introduced in ref. \citep{nair2023selective} to achieve this. This strategy allows us to independently and monotonically alter conserved quantities in a dynamical system based on the design of the forcing function \cite{Hanna20rbr}. In previous work \cite{nair2023selective}, this selective modification method has effectively stabilized a two-dimensional decaying turbulent flow by reshaping its energy and enstrophy cascade. In the present work, we apply different selective modification strategies to drive the two-dimensional Kolmogorov flow toward non-trivial steady-state solutions by influencing the time rate of change of energy and enstrophy within the turbulent flow. Our method is straightforward to implement and possesses an automatic mechanism for identifying forcing fields capable of altering turbulent flow dynamics. For a comprehensive understanding of the flow modification strategy, please refer to Section \ref{sec2}, while the results for the Kolmogorov flow are presented in Section \ref{sec3}, followed by concluding remarks in Section \ref{sec4}.

\section{Approach}
\label{sec2}
We consider the forced vorticity transport equation for Kolmogorov flow on a two-dimensional (2D) square bi-periodic domain $(x,y) \in [0, L] \times [0, L]$ as
\begin{equation}
\begin{split}
\frac{\partial \omega}{\partial t} &= \underbrace{ \frac{\partial \omega}{\partial y} \frac{\partial \psi}{\partial x} - \frac{\partial \omega}{\partial x} \frac{\partial \psi}{\partial y} }_{J(\omega,\psi)}+ \underbrace{ \frac{1}{Re} \nabla^2 \omega }_{d_{\omega} } \underbrace{ - n \cos(ny)}_{g_\omega}\\
& = N(\omega,\psi),
\vspace{-0.15in}
\label{eq1}
\end{split}
\end{equation}
where $\bm u = \nabla \times \left(\psi \hat{\bm k}\right)$ is the velocity field with stream function $\psi$, $\bm \omega = \nabla \times \bm u =  \omega \hat{\bm k}$ is the vorticity field, and $g_\omega = - n \cos(ny)$ is external forcing term. Here, the aggregate of non-linear contribution from convection terms is denoted by $J(\omega,\psi)$, and linear contribution from diffusion term is denoted by $d_\omega$. To simplify the notation, we denote the right-hand side by $N(\omega,\psi)$. Additionally, the mass conservation in an incompressible flow $\nabla \cdot \boldsymbol{u} = 0$ is satisfied by simultaneously solving 
\begin{equation}
    \nabla^2\psi = -\omega.
    \label{eq:continuity}
\end{equation}

Since the solution to Kolmogorov flow at large $n$ and $Re$ are unstable, the steady states only exist at the unstable equilibria when $N(\omega,\psi) = 0$. To achieve these equilibria, it is necessary to seek the roots of a set of non-linearly coupled equations, denoted as $N(\omega,\psi) = 0$, along with Eq.~\eqref{eq:continuity}. Solving this system can be computationally demanding and challenging, especially when using direct solution techniques like the Newton-Krylov hookstep method \citep{chandler2013invariant} or adjoint-based methods \citep{farazmand2016adjoint}. 

In the present study, we approach these equilibria solutions by solving the modified Kolmogorov flow equations defined as
\begin{equation}
    \begin{split}
        &\frac{\partial \omega}{\partial t} = N(\omega,\psi) + G(\omega,\psi),
    \end{split}
    \label{eq:controledKF}
\end{equation}
where a forcing term $G(\omega,\psi)$ is introduced. The design of the forcing term is crucial to ensure the stability of the solution to the modified Kolmogorov flow and drive the simulated flow of the modified Kolmogorov flow towards a steady state. In achieving this objective, we apply the concept of selective modification of inviscid invariants as discussed below. 

The flow kinetic energy $E(t)$ and enstrophy $\Omega(t)$ are defined as 
\begin{equation}
E(t) = \frac{1}{2 L^2} \int_{\mathcal{D}} \bm{u}\cdot\bm{u} \,\mathrm{d} {\bm{x}}, ~~~~~ \mathrm{and} ~~~~~ \Omega (t) = \frac{1}{2L^2} \int_{\mathcal{D}} \bm{\omega}\cdot\bm{\omega} \,\mathrm{d} {\bm{x}}.
\vspace{-0.05in}
\label{eq:energyEnstrophy_def}
\end{equation}
The above expressions for $E(t)$ and $\Omega (t)$ can be further simplified for the 2D periodic domain considered here as
\begin{equation}
E(t) = \frac{1}{2 L^2} \int_{\mathcal{D}} \psi \omega \,\mathrm{d} {\bm{x}}, ~~~~~ \mathrm{and} ~~~~~ \Omega (t) = \frac{1}{2L^2} \int_{\mathcal{D}} \omega^2 \,\mathrm{d} {\bm{x}},
\vspace{-0.05in}
\label{eq:energyEnstrophy_def}
\end{equation}
by applying integration by parts. 
%Furthermore, for a 2D inviscid flow over a periodic domain without any external forcing or control, $d_\omega = g_\omega = G(\omega,\psi) = 0$, $E(t)$ and $\Omega (t)$ can be shown to be invariant or conserved by integrating equation \eqref{eq1} multiplied by $\psi$ and $\omega$ respectively as follows:
%\begin{equation}
%\begin{split}
%    &\frac{\mathrm{d }E}{\mathrm{d}t} = \frac{1}{2L^2} \int_{\mathcal{D}} \psi \frac{\partial \omega}{\partial t} \,\mathrm{d} {\bm{x}} = \frac{1}{2L^2} \int_{\mathcal{D}} \psi J(\omega,\psi) \,\mathrm{d} {\bm{x}} = 0, \\ 
%    &\frac{\mathrm{d}\Omega}{\mathrm{d}t} =  \frac{1}{2L^2} \int_{\mathcal{D}} \omega \frac{\partial \omega}{\partial t} \,\mathrm{d} {\bm{x}} = \frac{1}{2L^2} \int_{\mathcal{D}} \omega J(\omega,\psi) \,\mathrm{d} {\bm{x}} = 0.
%%\vspace{-0.05in}
%\end{split}
%\label{eq:invariants}
%\end{equation}
%

Adopting the method of \citet{Hanna20rbr} in the context of inviscid flow without non-conservative forces of viscous $d_\omega$ and body force terms $g_\omega$: for $m~ (\leq n)$ independent conserved quantity $Q_i(\bm{\mathrm{x}})$ such that $\dot{\bm{\mathrm{x}}} = J(\bm{\mathrm{x}}) + G(\bm{\mathrm{x}}), x \in \mathbb{R}^n$, $\dot{Q}_i$ can be written as
\begin{equation}
    \dot{Q}_i = \frac{d Q_i}{d \bm{\mathrm{x}}} \cdot \dot{\bm{\mathrm{x}}} = \frac{d Q_i}{d \bm{\mathrm{x}}} \cdot \bigg[J(\bm{\mathrm{x}}) + G(\bm{\mathrm{x}})\bigg] = \frac{d Q_i}{d \bm{\mathrm{x}}} \cdot G(\bm{\mathrm{x}}).
    \label{eq:conservedEqn}
\end{equation}
Here, the term $\frac{d Q_i}{d \bm{\mathrm{x}}} \cdot J(\bm{\mathrm{x}}) = 0$ by definition and the quantities $Q_i$ are conserved in absence of any forcing term $G(\bm{\mathrm{x}})$. According to the proportional dissipation formalism \citep{Hanna20rbr,MatteoHanna21rbr}, $G(\bm{\mathrm{x}})$ can be designed as 
\begin{equation}
    G(\bm{\mathrm{x}}) = -\frac{1}{2}\bm{B}\mathrm{adj}(\bm{B}^T\bm{B})\bm{\epsilon}
\end{equation}
that satisfies equation \eqref{eq:conservedEqn} and can modify conserved quantities independently by tuning individual values of $\epsilon_i$. Here matrix $\bm{B} = \{b_1,b_2,\dots,b_m\}$ represents $m$ independent $b_i = \frac{d Q_i}{d \bm{\mathrm{x}}}$. In the present case with $Q_i = (E,\Omega)$ and $B = (\psi,\omega)$, $G(\omega,\psi)$ can be computed as 
\begin{equation}
    G(\omega,\psi) = -\frac{1}{2}\begin{bmatrix}
           \psi \\
           \omega
         \end{bmatrix}
         \mathrm{adj}\left(
         \begin{bmatrix}
            \psi^2 & \psi\omega\\
            \psi\omega & \omega^2
        \end{bmatrix}\right)
        \begin{bmatrix}
           \epsilon_1 \\
           \epsilon_2
         \end{bmatrix},
    \label{eq:controlFunction1}
\end{equation}
where $\mathrm{adj}(\bullet)$ denotes the adjugate (transpose of the cofactor matrix). Simplifying Eq~\eqref{eq:controlFunction1} we get
\begin{equation}
\begin{split}
G(\omega,\psi) = &-\frac{\epsilon_1}{2}\left( \psi \int_{\mathcal{D}} \omega^2\,\mathrm{d} {\bm{x}} - \omega \int_{\mathcal{D}} \psi \omega\,\mathrm{d} {\bm{x}}\right)\\ &- \frac{\epsilon_2}{2}\left( \omega \int_{\mathcal{D}} \psi^2 \mathrm{d}{\bm{x}} - \psi \int_{\mathcal{D}} \psi \omega\,\mathrm{d} {\bm{x}} \right).
\vspace{-0.05in}
\label{eq:controlFunction2}
\end{split}
\end{equation}

Incorporating the non-conservative governing equations of Kolmogorov flow, which account for viscous and body force terms, we can express the domain-integrated instantaneous energy dissipation rate, denoted as $D(t)$, and the energy input rate, denoted as $I(t)$, as follows:
\begin{equation}
\begin{split}
&D(t) = -\frac{1}{2L^2} \int_{\mathcal{D}} b_1d_\omega\,\mathrm{d} {\bm{x}} = \frac{1}{2L^2 Re} \int_{\mathcal{D}} \omega^2\,\mathrm{d} {\bm{x}}, \\
&I(t) = \frac{1}{2L^2} \int_{\mathcal{D}} b_1 g_\omega\,\mathrm{d} {\bm{x}}.
\vspace{-0.05in}
\label{eq4}
\end{split}
\end{equation}
These equations reveal that the energy dissipation rate is directly proportional to the enstrophy of the system. Consequently, in viscous flows, the energy and enstrophy are inherently coupled and can only be completely decoupled in inviscid flows. However, in practical applications involving turbulent flows at high Reynolds numbers ($Re$), the viscous contributions become relatively small. This allows for the effective use of selective modification methods, as demonstrated in the subsequent section. 

One can also define the contribution of $G(\omega,\psi)$ to the time rate of change of energy $F_1$ and enstrophy $F_2$ as
\begin{equation}
\begin{split}
&F_{1}(t) = \frac{1}{2L^2}\int_{\mathcal{D}} b_{1} G(\omega,\psi) \,\mathrm{d} {\bm{x}},\\
&F_{2}(t) = \frac{1}{2L^2}\int_{\mathcal{D}} b_{2} G(\omega,\psi) \,\mathrm{d} {\bm{x}}, 
\vspace{-0.05in}
\label{eq5}
\end{split}
\end{equation}
which can monotonically drive the 2D Kolmogorov flow to an equilibrium solution
\begin{equation}
\frac{\mathrm{d }E}{\mathrm{d}t} = 0, ~~~~\frac{\mathrm{d }\Omega}{\mathrm{d}t} = 0. 
\label{eq6}
\vspace{-0.05in}
\end{equation}

To use the forcing function defined by Eq.~\eqref{eq:controlFunction2}, one also needs to specify the coefficients $\epsilon_1$ and $\epsilon_2$. For an inviscid case without external forcing $d_\omega = g_\omega = 0$, considered in \citet{nair2023selective}, the following behaviors are observed: (\emph{a}) when $\epsilon_1 = 0$, energy remains conserved. Enstrophy increases if $\epsilon_2 < 0$ or decreases if $\epsilon_2 > 0$.
(\emph{b}) Conversely, when $\epsilon_2 = 0$, enstrophy remains conserved. Energy increases if $\epsilon_1 < 0$ or decreases if $\epsilon_1 > 0$. This implies that the first term of $G(\omega,\psi)$ alters the time rate of change of energy such that $\frac{\mathrm{d }E}{\mathrm{d}t} \rightarrow 0$ in finite time, while the second term modifies enstrophy in such a way that $\frac{\mathrm{d }\Omega}{\mathrm{d}t} \rightarrow 0$ in finite time.

With the forcing defined by Eq.~\eqref{eq:controlFunction2}, the contribution of $G(\omega,\psi)$ to the time rate of change of energy and enstrophy is given by
\begin{equation}
\begin{split}
F_{1,2} = &\int_{\mathcal{D}} b_{1,2} G(\omega,\psi) \,\mathrm{d} {\bm{x}}\\
= &-\epsilon_{1,2}\left(\ \int_{\mathcal{D}} \psi^2 \,\mathrm{d} {\bm{x}} \int_{\mathcal{D}} \omega^2 \,\mathrm{d} {\bm{x}}\right) \sin^2(\beta), % \equiv -\epsilon_i F,
\vspace{-0.05in}
\label{eq13}
\end{split}
\end{equation}
where $\beta$ indicates the cosine similarity or ``angle'' between the vectors $\psi$ and $\omega$ in the function space, defined by $\cos{\beta} = \int_\mathcal{D} \psi \omega \,\mathrm{d} \bm{x} / \sqrt{\int_\mathcal{D} \psi^2 \,\mathrm{d} \bm{x} \int_\mathcal{D} \omega^2 \,\mathrm{d} \bm{x}}$. This essentially means that the forcing rates vanish when $\beta=0$, i.e., when $\psi$ and $\omega$ ``align'', and the quantity $E/\epsilon_1  - \Omega/\epsilon_2$ is conserved \citep{nair2023selective}.

It's important to highlight that the forcing term derived in Eq.~\eqref{eq:controlFunction2} in the modified Kolmogorov flow does not correspond to the steady state solution of the original governing equations for the Kolmogorov flow, as described in Eq.~\eqref{eq1}. This distinction is demonstrated through a series of numerical experiments in Section \ref{sec3}, where the Kolmogorov flow evolves without a forcing term but is initialized with the steady solution of the modified Kolmogorov flow. The trajectory of the flow evolution and the flow's structure illustrate the effectiveness of the proposed method in stabilizing the Kolmogorov flow and providing a close initial guess for the search of non-trivial roots in the original governing equations.

\section{Results and Discussion}
\label{sec3}

\begin{figure*}
\centering
\includegraphics[width=0.95\textwidth]{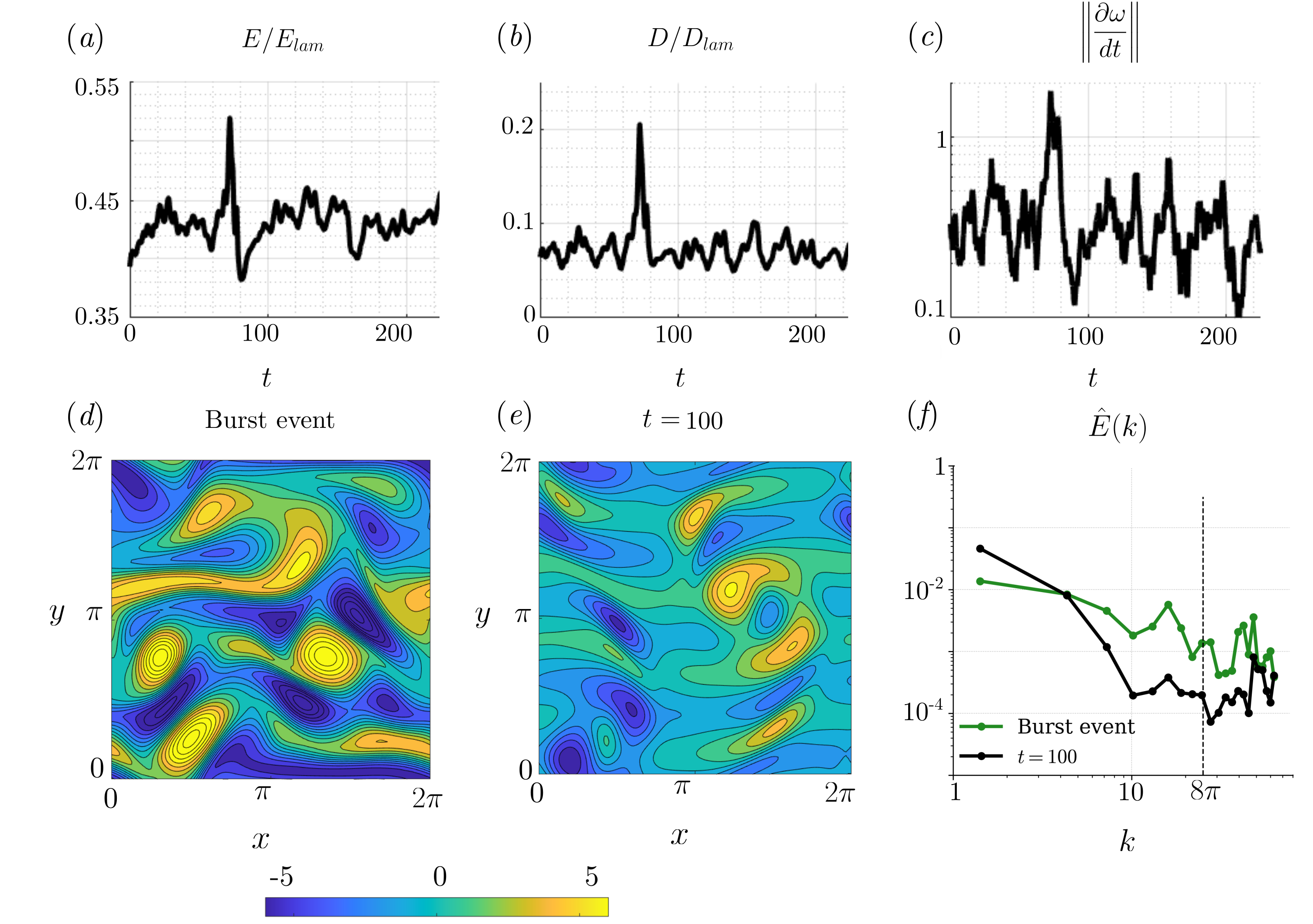}
\caption{Time history of (\emph{a}) normalized energy ($E/E_{lam}$), (\emph{b}) normalized energy dissipation rate ($D/D_{lam}$), (\emph{c}) residual of the right-hand side of the governing equation $\left(\|\frac{\partial \omega}{\partial t}\|\right)$. (\emph{d}) Vorticity snapshot at the burst event, (\emph{e}) vorticity snapshot at $t = 100$ and (\emph{f}) energy spectra at the burst event and fully developed Kolmogorov flow ($t = 100$) from simulation.}
\label{fig1}
\end{figure*}

2D direct numerical simulations of Kolmogorov flow are performed using a Fourier spectral method and a fourth-order Runge--Kutta time integration scheme on a bi-periodic computational domain ($L = 2\pi$) with $128 \times 128$ grid points in the $x$- and $y$-directions \citep{taira2016network}. The laminar equilibrium solution for Kolmogorov flow is characterized by the following values:
\begin{equation}
E_\text{lam} = \frac{Re^2}{4n^4}, ~~~ D_\text{lam} = I_\text{lam} = \frac{Re}{2n^2}
\end{equation}
In the present study, we fix $Re = 40$ and $n = 4$ as in previous studies \cite{chandler2013invariant,farazmand2016adjoint} for which the solution is unstable.

The time history of energy and its dissipation rate (which is proportional to enstrophy, see Eq.~\ref{eq4}) for Kolmogorov flow are shown in Figures \ref{fig1}(\emph{a} and \emph{b}), respectively. We see intermittent bursts of oscillations in both the time series, which is ubiquitous in turbulent flows \citep{batchelor1949nature}. The burst in energy is a consequence of the alignment of flow velocity with external forcing $g_\omega$ \citep{farazmand2016adjoint}. The trajectories return back to the nominal oscillatory state due to a subsequent increase in the energy dissipation rate. We see in Figure~\ref{fig1}(\emph{c}) that the $L_2$ norm of the $\partial\omega/\partial t$ or $F(\omega,\psi)$  jumps to a value greater than $\mathcal{O}(1)$ with intermittent bursts. Henceforth, we refer to this norm as the residual. The lower limit of the residual is at $\mathcal{O}(10^{-1})$. The vorticity fields at a burst event and a nominal oscillatory state ($t = 100$) are shown in Figures \ref{fig1}(\emph{d} and \emph{e}). Figure~\ref{fig1}(\emph{f}) shows the isotropic energy spectra of the flow for the two states shown in Figures\ref{fig1}(\emph{d} and \emph{e}). The isotropic energy spectrum for 2D turbulence is define as $\hat{E}(k) = \pi k \langle|\hat{\bm{u}}(\bm{k})|^2\rangle$ where the average $\langle\cdot\rangle$ is computed over all $|\bm{k}| = k$ and $\hat{\bm{u}}(\bm{k}) = \int_{\mathcal{D}} \bm{u}(\bm{x})e^{i\bm{k}\cdot\bm{x}}d\bm{x}$ ~\citep{boffetta2012two}. Observing the isotropic energy spectra, it becomes evident that during a burst event, a greater amount of energy accumulates near the wave number $k = k_f$ (where $k_f = 8\pi$ represents the forcing wave number). This energy subsequently redistributes itself towards lower wave numbers, primarily because of the phenomenon known as inverse energy cascading, which is a characteristic feature of fully developed Kolmogorov flow, as described in \citet{shu1993effective}.

\begin{figure*}
\centering
\includegraphics[width=0.85\textwidth]{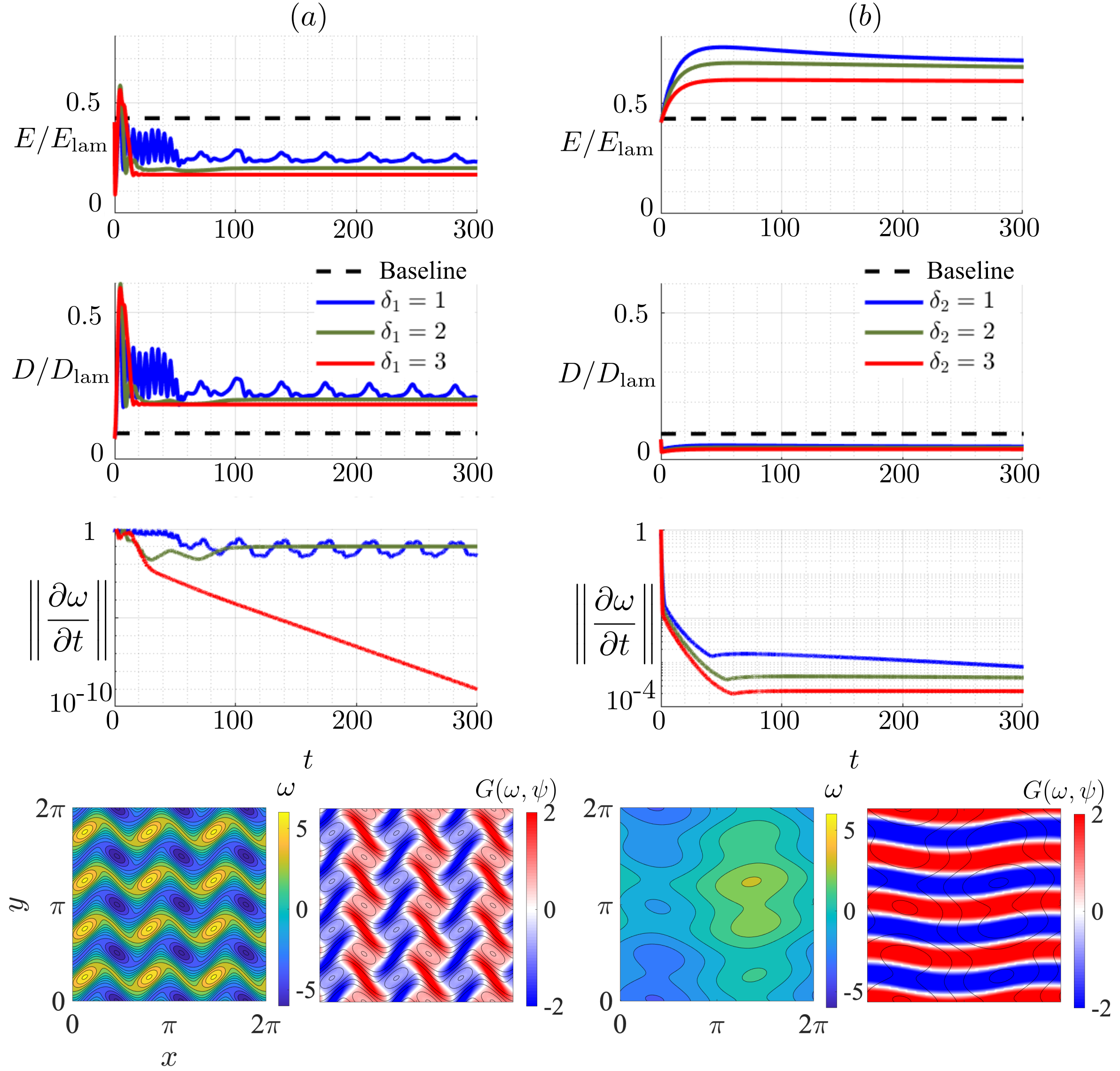}
\caption{Time history of normalized energy, normalized energy dissipation rate and residual for selective modification of 2D Kolmogorov flow at $Re = 40$: Modification of (\emph{a}) energy ($\delta_2=0, \delta_1>0$) and (\emph{b}) enstrophy ($\delta_1 = 0, \delta_2>0$) for different rate parameters. The mean quantities from the Kolmogorov flow are shown with black dashed lines as a baseline for comparison. Also shown at the bottom are vorticity snapshots and forcing fields $G(\omega,\psi)$ superimposed on vorticity contours (black) at the steady state solution ($t = 300$) of modified flows corresponding to $\delta_{1,2} = 3$.}
\label{fig2}
\end{figure*}

We use the solution at $t = 100$ shown in Figure~\ref{fig1}(\emph{e}) as the initial condition to simulate the modified Kolmogorov flow. To use $G(\omega,\psi)$ as a forcing term to the flow, we need to specify the energy and enstrophy modification rates in Eq.~\eqref{eq:controlFunction2} through $\epsilon_1$ and $\epsilon_2$. We define $\epsilon_1 = \delta_1 E_0/F_0$ and $\epsilon_2 = \delta_2 \Omega_0/F_0$ such that energy and enstrophy modifying components of the forcing have magnitudes similar to the initial energy $E_0$ and initial enstrophy $\Omega_0$ of the system, respectively \citep{nair2023selective}. Positive and negative values of $\delta_{1,2}$ represent extraction and injection of requisite quantity through selective modification. Here, $F_0 \equiv -\tfrac{F_{1,2}}{\epsilon_{1,2}}$ (see Eq.~\ref{eq13} for $F_{1,2}$) and a higher magnitude of $\delta_{1,2}$ implies faster modification of the requisite quantities. Figure~\ref{fig2}(\emph{a} and \emph{b}) show the effect of selectively modifying the energy ($\delta_2 = 0$) and enstrophy ($\delta_1 = 0$) of the flow, respectively. The time history of energy, enstrophy, and residual of the vorticity transport equation are presented for both flow modification strategies. We choose three different normalized forcing rates $\delta_{1,2} = \{1,2,3\}$ to show the convergence of the obtained solutions. The mean quantities from Kolmogorov flow are shown in black dashed lines for a baseline comparison. 

For selective energy modification in Figure~\ref{fig2}(\emph{a}), after an initial transient, the energy decreases to a lower value relative to the baseline mean, while the energy dissipation rate relatively increases. Here, the increase in energy dissipation rate is due to the presence of non-conservative forces in the governing equation. For $\delta_1 = 1$, oscillations in the flow persist, while steady state trajectories are reached for higher rates. The residual of $\delta_1 = 2$ solution saturates at $\mathcal{O}(10^{-1})$, while that of $\delta_1 = 3$ attenuates to $\mathcal{O}(10^{-10})$ at $t = 300$. The vorticity field $\omega$ and the forcing term $G(\omega,\psi)$ to maintain this steady state of the flow is also shown in Figure~\ref{fig2}(\emph{a}). We see the presence of horizontal layers of alternating and connected positive and negative vorticity fields. The negative vortex structures and connecting layers between the positive vortex structures experience positive forcing fields and vice versa.

For selective enstrophy modification in Figure~\ref{fig2}(\emph{b}), the trajectories quickly saturate to a steady state value except for the case with $\delta_2 = 1$. The energy dissipation rate or enstrophy drops to a similar lower value for all cases compared to the baseline mean, while the energy increases to a relatively higher value. The residual of the $\delta_2 = 3$ forcing input saturate around $\mathcal{O}(10^{-4})$. Increasing the forcing rate further doesn't have any significant effect on the steady-state energy dissipation rate or the residual. The vorticity field at the steady state consists of large-scale vortical structures and the corresponding forcing field consists of sheets of positive and negative magnitude similar to the sinusoidal forcing term $g_\omega$. 

\begin{figure*}
\centering
\includegraphics[width=0.9\textwidth]{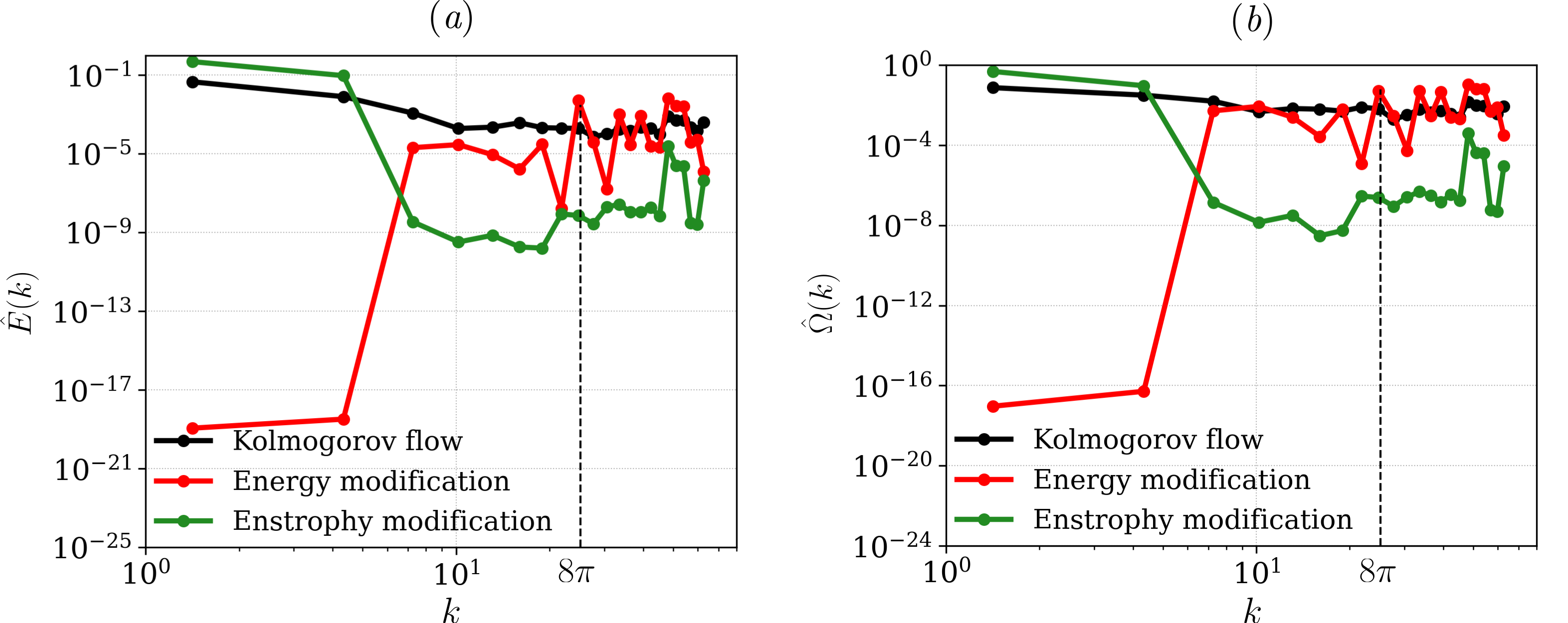}
\caption{(\emph{a}) Energy and (\emph{b}) enstrophy spectra comparison between Kolmogorov flow solution at $t = 100$ (initial condition to the modified Kolmogorov flow simulations) and steady-state solutions of modified Kolmogorov flow using forcing functions that modify energy ($\delta_1 = 3$) and enstrophy ($\delta_2 =3$).}
\label{fig2_spectra}
\end{figure*}

Figure~\ref{fig2_spectra}(\emph{a} and \emph{b}) present comparisons of isotropic energy and enstrophy spectra, respectively of the steady-state solutions obtained with energy and enstrophy modifications in the Kolmogorov flow. Energy modification of the Kolmogorov flow has damped the energy-containing large-scale structures at low wave numbers while preserving the scales with large wave numbers and an opposite pattern is seen for the case with enstrophy modification. Moreover, the energy and enstrophy spectra show that the enstrophy modification of the Kolmogorov flow has damped the scales close to energy injection wave number ($k_f = 8\pi$) which is in line with the fact that $G(\omega,\psi)$ partially negates the contribution of $g_\omega$ in Figure~\ref{fig2}(\emph{b}).

\begin{figure*}
\centering
\includegraphics[width=\textwidth]{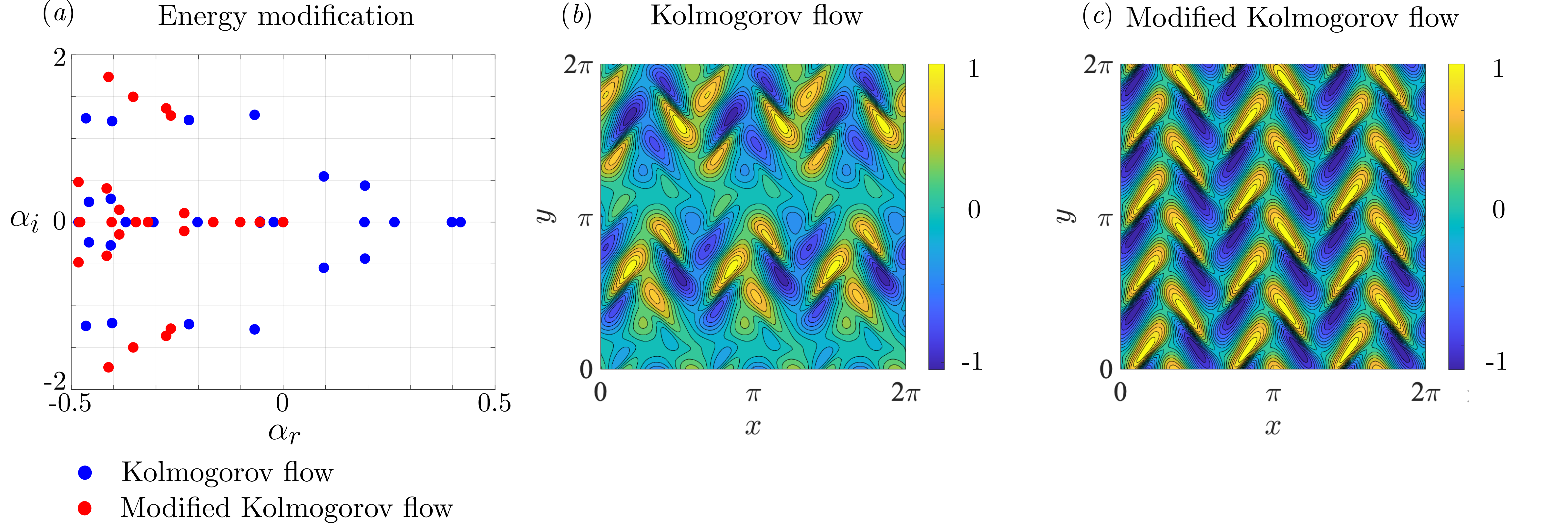}
\caption{(\emph{a}) Eigenvalue spectra from the stability analysis of Kolmogorov and modified Kolmogorov flow, (\emph{b}) eigenvector corresponding to the most unstable eigenvalue of Kolmogorov flow, (\emph{c}) eigenvector corresponding to the eigenvalue of modified Kolmogorov flow with largest growth rate.}
\label{eigenValues}
\end{figure*}

To understand the stability properties of the steady-state solution obtained from energy forcing shown in Figure~\ref{fig2}(\emph{a}), we perform a linear stability analysis of the Kolmogorov flow about this modified steady-state. The stability of the solution is examined using eigenvalues of the linearized governing equations in the presence and absence of steady-state external forcing $G(\omega,\psi)$. To construct the linear operator, we introduce an infinitesimal impulse perturbation at each grid location $\boldsymbol{x}_j$ and measure the change in vorticity evolution at $\boldsymbol{x}_i$. The linear operator is computed as 
\begin{equation}
\begin{split}
A_{ij} = &[N(\omega + \epsilon \delta (\boldsymbol{x}_j)) - N(\omega)]_{x_i}/\epsilon\\
+ &[G(\omega + \epsilon \delta (\boldsymbol{x}_j)) - G(\omega)]_{x_i}/\epsilon,
\label{linop}
\end{split}
\end{equation}
where the right-hand side of Eq. (\ref{eq:controledKF}) is used with $\epsilon = 0.001$. Here $\omega$ is the modified steady-state solution. In the absence of the selective modification forcing, the second term in Eq.~(\ref{linop}) vanishes. The detailed method of calculating the eigenvalues and eigenvector with the above linear operator can be found in ref.\cite{yeh2021network}. Some of the leading eigenvalues ($\alpha_r + i\alpha_i$) about the modified steady state in the presence and absence of selective forcing are shown in Figure~\ref{eigenValues}(\emph{a}). We observe that in the presence of forcing, all the eigenvalues are damped with non-positive growth rates ($\alpha_r \leq 0$), as indicated by the red dots. In the absence of selective forcing input, multiple eigenvalues have positive growth rates ($\alpha_r >0$), as indicated by the blue dots. We also show the eigenvector with maximum growth rate in Figure~\ref{eigenValues}(\emph{b}) which eventually causes the flow to become unstable and leave the attractor of the equilibrium solution. In Figure~\ref{eigenValues}(\emph{c}), we show the dominant damped eigenvector in the presence of selective energy forcing. 

\begin{figure*}
\centering
\includegraphics[width=\textwidth]{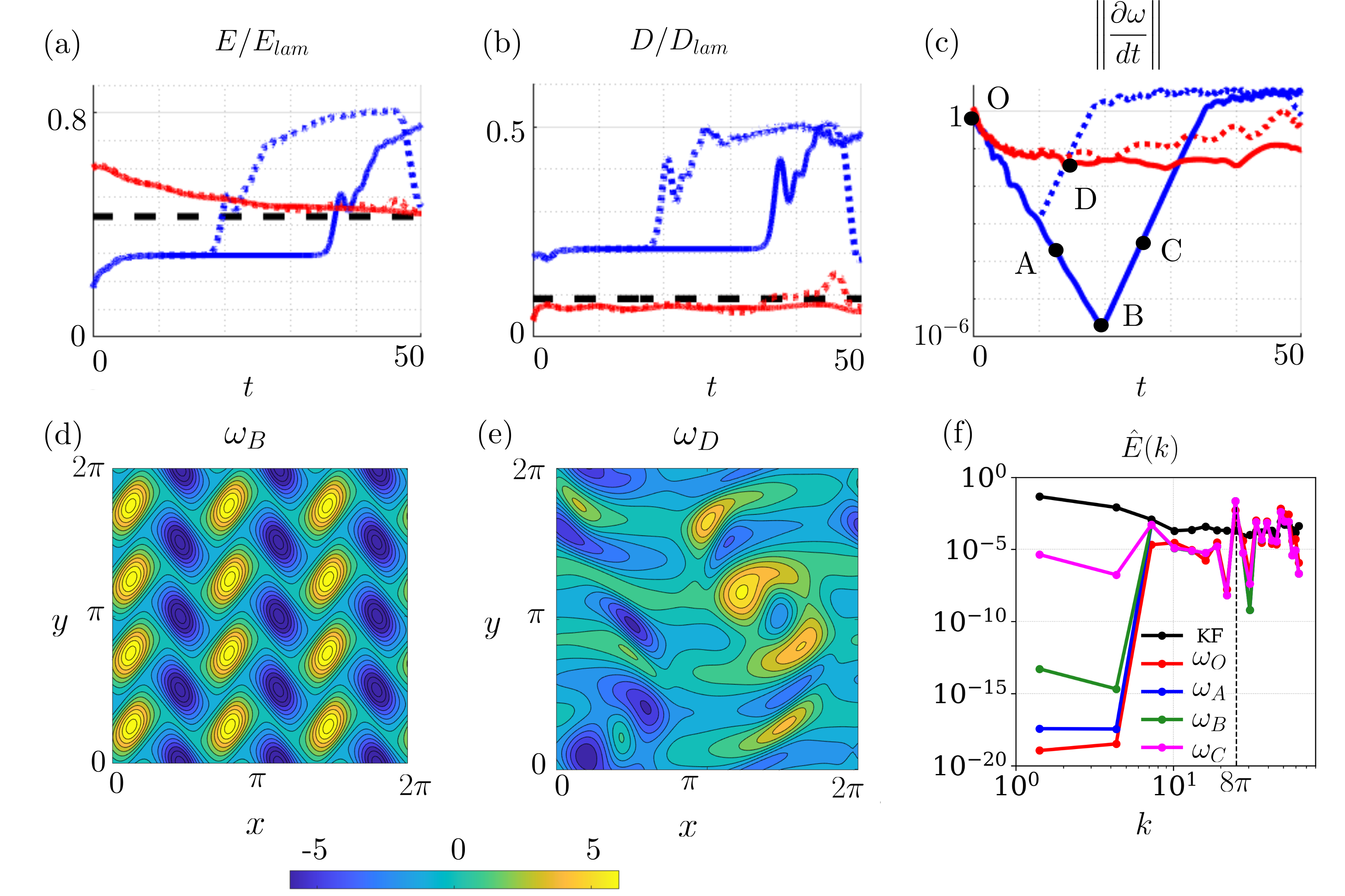}
\caption{Time history of (\emph{a}) normalized energy, (\emph{b}) normalized energy dissipation rate, and (\emph{c}) residual for the Kolmogorov flow released from the modified steady-state with $G(\omega,\psi) = 0$. The blue and red lines indicate time histories starting from modified energy and enstrophy steady states, respectively. The dotted lines correspond to the flow released from modified steady states at $t = 150$. The mean quantities from the Kolmogorov flow are shown with black dashed lines as a baseline for comparison. (\emph{d}, \emph{e}) The vorticity field corresponds to the states marked B and D in Fig (\emph{c}). (\emph{f}) Spectra comparison of Kolmogorov flow (KF) as the baseline, $\omega_O$ (steady state solution of modified Kolmogorov flow with energy modification), $\omega_A$, $\omega_B$, and $\omega_C$ are states shown in Figure (\emph{c}) as Kolmogorov flow solution evolves with from initial condition without forcing.}
\label{fig3}
\end{figure*}

The nature of Kolmogorov flow around the steady-state solution obtained from the modified Kolmogorov flow can also be seen by observing the evolution of flow from this steady-state as initial condition and switching off the external forcing, \emph{i.e.}, $G(\omega,\psi) = 0$. The time history of energy and energy dissipation rate from these simulations are shown in Figures \ref{fig3}(\emph{a} and \emph{b}), respectively. The time history of the flow released from the modified energy steady state (vorticity field in the bottom panel of Figure~\ref{fig2} (\emph{a})) is shown in blue. The energy of the flow quickly saturates to $E/E_\text{lam} = 0.2917$ with corresponding $I/I_\text{lam} = D/D_\text{lam} = 0.2098$ and then jumps out of this state at a later time with growth in perturbations as all invariant solutions are unstable. To test that the growth in perturbations causes the flow to deviate from the saturated state, we release the flow from an earlier modified energy state at $t = 150$ with a residual of $\mathcal{O}(10^{-6})$. The time histories for this alternate initial condition are shown with the dotted blue lines. Although the flow saturates at the same energy level, growing perturbations cause the flow to deviate much earlier in this case. For the flow released from the modified enstrophy steady state (vorticity field in the bottom panel of Figure~\ref{fig2}(\emph{b})), the time histories (shown in red in Figure~\ref{fig3}) return to the mean of the baseline of Kolmogorov flow solution. 

\begin{figure*}
\centering
\includegraphics[width=\textwidth]{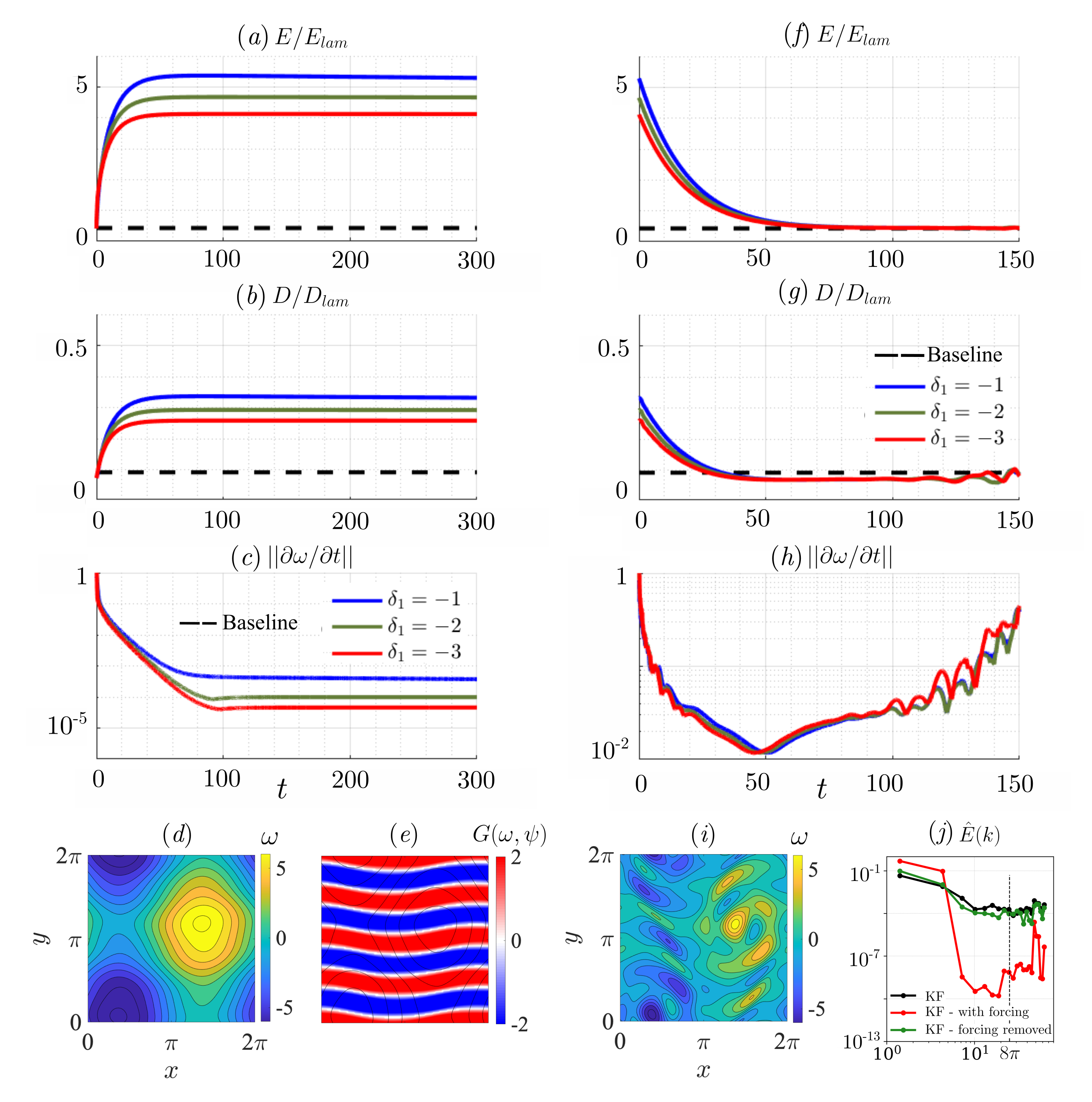}
\caption{Time history of (\emph{a}) normalized energy, (\emph{b}) normalized energy dissipation rate, and (\emph{c}) residual for selective energy modification of 2D Kolmogorov flow ($\delta_2=0, \delta_1<0$). Shown at the bottom are (\emph{d}) vorticity snapshots and (\emph{e}) forcing fields $G(\omega,\psi)$ superimposed on vorticity contours (black) at the equilibrium solution ($t = 300$) of the modified flow corresponding to $\delta_1 = -3$. Time history of (\emph{f}) normalized energy, (\emph{g}) normalized energy dissipation rate, and (\emph{h}) residual of Kolmogorov flow released from modified energy steady state with $G(\omega,\psi) = 0$. (\emph{i}) Vorticity field corresponding to a residual of $\mathcal{O}(10^{-2})$ for flow released from modified energy steady state and (\emph{j}) comparison of energy spectra. Here, KF means the Kolmogorov flow, KF with forcing means modified Kolmogorov flow with forcing for energy modification and KF forcing removed means Kolmogorov flow solution initialized from steady state of modified Kolmogorov flow and forcing switched off.}
\label{fig4}
\end{figure*}

The residual for the flow released from the modified energy steady state with $G(\omega,\psi) = 0$ reaches nearly $\mathcal{O}(10^{-6})$, as seen in Figure~\ref{fig3}(\emph{c}), indicating that this is also close to a true equilibrium solution of the Kolmogorov flow. The corresponding vorticity field is shown in Figure~\ref{fig3}(\emph{d}) which contains patterns similar to the energy-modified steady-state, although with isolated vortex structures. This equilibrium solution for Kolmogorov flow has earlier been reported by \citet{farazmand2016adjoint} (refer to equilibrium $E_6$ in table 1). 

The residual for the flow released from the modified enstrophy steady state only reduces to $\mathcal{O}(10^{-2})$ with a vorticity snapshot shown in Figure~\ref{fig3}(\emph{e}). No inflection in the residual is observed indicating that a true equilibrium is not found in this case. Although not shown, based on our tests, increasing the enstrophy modification rate up to $\delta_2 = 20$, we get a residual of $\mathcal{O}(10^{-6})$ with the application of a forcing input. However, releasing the flow from this modified steady state does not yield any near equilibrium solution either, and the time histories are similar to the red lines in Figures \ref{fig3}(\emph{a} -- \emph{c}). Such behavior of the present flow modification strategy may be a consequence of the inverse energy cascade in Kolmogorov flow. Since the energy flows towards smaller wave numbers from $k_f$, energy extraction at low wave numbers stabilizes the flow near a non-trivial unstable equilibrium of the Kolmogorov flow. However, enstrophy modification only influences scales with moderate to larger wave numbers and fails to reach any equilibrium solution of the Kolmogorov flow. 

Figure~\ref{fig3}(\emph{f}) shows the evolution of spectral scales in this simulation with forcing switched off. For reference, isotropic energy spectra from a nominal Kolmogorov flow solution are shown with a black line. The events from the simulation are marked with letters O (steady state solution of modified Kolmogorov flow as an initial condition), A (solution before minimum residual is reached), B (solution with minimum residual), and C (solution after the flow moves away from the equilibrium or the minimum residual point) in Figure~\ref{fig3}(\emph{c}) and the corresponding isotropic energy spectra are shown in Figure  \ref{fig3}(\emph{f}). It is observed that the energy at lower wave numbers increases with the flow evolution due to the inverse energy cascade in Kolmogorov flow. A small change in spectra is observed near the plateau region that corresponds to the interesting scales in the non-trivial solution of Kolmogorov flow as shown in Figure~\ref{fig3}(\emph{d}) indicating that the steady state solution of the modified Kolmogorov flow is very similar to the equilibrium solution of Kolmogorov flow. The small differences are due to the finite values of $G(\omega,\psi)$ at the steady state of modified Kolmogorov flow which quickly vanished for the spectra of states A, B, and C after the forcing is switched off.

\begin{figure*}
\centering
\includegraphics[width=0.8\textwidth]{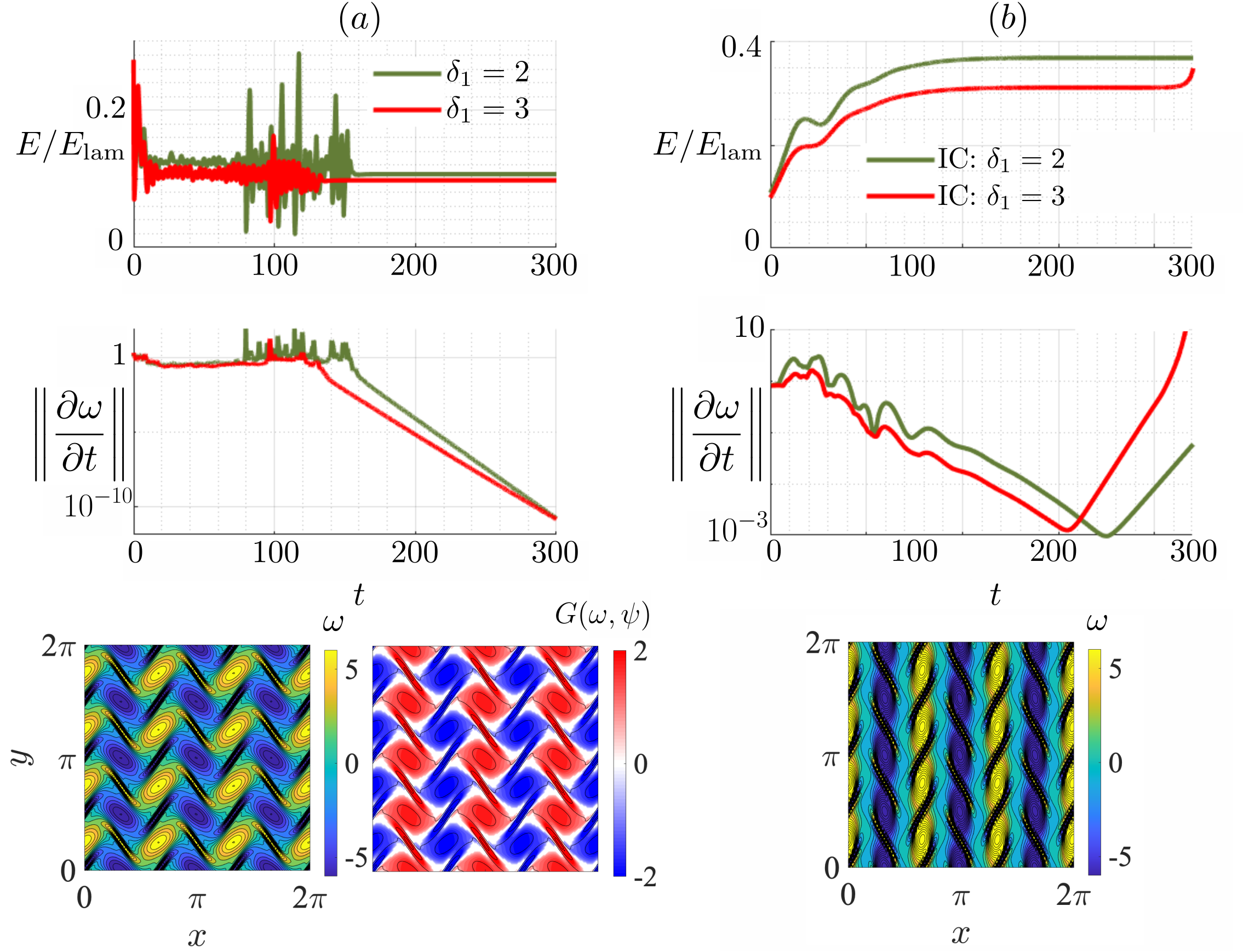}
\caption{Normalized energy and residual time histories for 2D Kolmogorov flow at $Re = 60$ with (\emph{a}) selective energy modification  ($\delta_2=0, \delta_1>0$) and (\emph{b}) releasing the flow from modified energy steady state. The vorticity and the forcing fields for the modified energy steady state ($\delta_1 = 2$) are shown on the bottom left. On the bottom right is the vorticity field corresponding to a residual of $\mathcal{O}(10^{-3})$ for flow released from the modified energy steady state.}
\label{fig5}
\end{figure*}

\begin{figure*}
\centering
\includegraphics[width=0.8\textwidth]{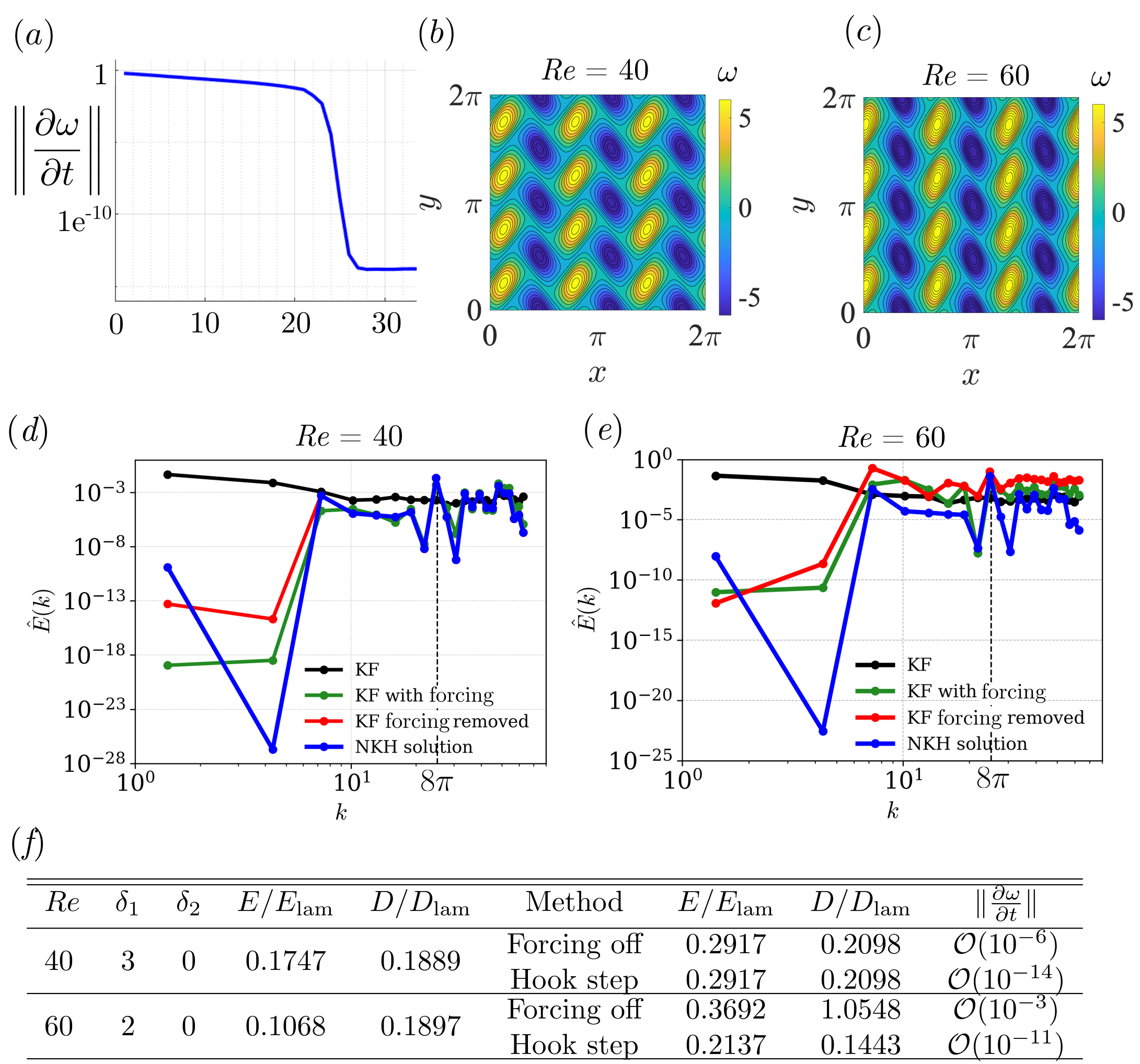}
\caption{Newton-GMRES hook step method with the modified energy steady state as the initial guess: (\emph{a}) Convergence of residual at $Re = 40$, true equilibrium vorticity snapshot for (\emph{b}) $Re = 40$ and (\emph{c}) $Re = 60$, spectra comparison for (\emph{d}) $Re = 40$ and (\emph{e}) $Re = 60$. KF means Kolmogorov flow, KF with forcing means modified Kolmogorov flow with forcing for energy modification, KF forcing removed means Kolmogorov flow solution initialized from steady state of modified Kolmogorov flow and forcing switched off and NKH solution means solution of $F(\omega,\psi) = 0$ obtained using Newton-Krylov hook step method and modified energy steady state as initial condition. (\emph{f}) Table summarizing solutions at modified steady states and subsequent solutions with forcing off and Newton-Krylov hook step method.}
\label{fig6}
\end{figure*}

Now, instead of selectively decreasing the energy of the flow, we increase it by setting $\delta_1 < 0$. The time history and flow fields for the modified flow are shown in Figures~\ref{fig4}(\emph{a} -- \emph{c}). Both energy and energy dissipation rates in this case increase compared to the mean baseline value. The flow with forcing input rate of $\delta_1 = -3$ has a residual of about $\mathcal{O}(10^{-5})$. The corresponding vorticity and forcing at the steady state ($t = 300$) are shown in Figures~\ref{fig4}(\emph{d}) and~\ref{fig4}(\emph{e}), respectively. The vorticity field is similar to that obtained in two-dimensional decaying turbulent flows of two strong vortices of opposite signs, where all the energy is accumulated at the lowest wavenumber \citep{nair2023selective}. The vortical structures and the forcing field for this modified steady state also look similar to that of the decreasing dissipation rate solution in Figure~\ref{fig2}(\emph{b}).

The flow released from the above-modified energy steady state with $G(\omega,\psi) = 0$ is shown in Figure~\ref{fig4}(\emph{f}--\emph{h}). The energy and dissipation rate decreases exponentially and saturates for a brief period of time for all cases with $E/E_\text{lam} = 0.6854$ and $I/I_\text{lam} = D/D_\text{lam} = 0.0704$. The residual drops to $\mathcal{O}(10^{-2})$ before perturbation growth causes the solution to deviate from the saturated state. The vorticity field of the flow solution around $t = 50$ is shown in Figure~\ref{fig4}(\emph{i}). This solution has earlier been observed by \citet{chandler2013invariant} (see Figure 7(a) for more details). This is a state that bifurcates off the basic solution for Kolmogorov flows beyond $Re = 15$. The energy spectra shown in Figure~\ref{fig4}(\emph{j}) resembles the selective enstrophy decrease case seen earlier in Figure~\ref{fig2_spectra}(\emph{a}) and instabilities set in quickly to drive the stabilized flow back to unsteadiness. This is evident from the close resemblance of the Kolmogorov flow spectra after forcing was removed from the baseline Kolmogorov flow. In the Kolmogorov flow with forcing, there is a distinct accumulation of energy at the lower wavenumbers. The energy increase at low wave numbers is due to accumulation of energy from the higher wavenumbers and the additional energy introduced into the flow through non-conservative forcing. As the steady-state selective energy forcing $G(\omega, \psi)$ approximately counterbalances the external forcing term  $g_\omega$ (see resemblance of $G(\omega,\psi)$ in Figure~\ref{fig4}(\emph{e}) to $g_\omega$ in the governing equation), the non-conservative components of energy are small and the steady-state structures resemble the dipolar structures from the decaying isotropic turbulence \citep{smith1993bose, nair2023selective}. 

To show the effect of Reynolds number on the obtained steady-state solutions of modified Kolmogorov flow, we perform the selective energy modification of Kolmogorov flow at $Re = 60$ in Figure~\ref{fig5} (\emph{a}) with $\delta_1 = \{2,3\}$. We see a lot of fluctuations in the initial transients of the energy of the flow for both forcing rates. After this period, the flow settles to a steady state with a residual of $\mathcal{O}(10^{-10})$. The corresponding vorticity field has similar patterns to that of the modified energy steady state at $Re = 40$, except for the presence of strong elongated vortical structures in the connecting layers. The magnitude of the input force is also high at these connecting structures, as is evident from the forcing field. Once the flow is released from this modified steady state with the forcing turned off $G(\omega,\psi) = 0$, the energy rises and saturates at the solution with a residual of $\mathcal{O}(10^{-3})$ as seen in Figure~\ref{fig5}(\emph{b}). The corresponding vorticity field is shown in the bottom panel which contains some elongated filaments of opposite vorticity sign adjacent to the strong vortex cores. 

We note that the trajectory to the modified steady states of the flow obtained in this work is dependent on the forcing rates and the initial condition of the baseline flow. Although not shown, we have tested different initial conditions of the baseline flow and obtained very similar modified steady-state solutions. The main difference originating in simulations from different initial conditions is the magnitude of the coefficients ($\delta_1$ or $\delta_2$) needed to direct the flows to this modified state. \emph{e.g.}, for the initial condition near a burst event, a higher forcing rate $\delta_1 = 6$ is needed to reach the same steady state earlier shown in Figure~\ref{fig2}.

To assess the usefulness of the steady states obtained from modified Kolmogorov flow simulations in finding the equilibrium solutions of Kolmogorov flow, the vorticity field corresponding to the modified energy steady state in Figure~\ref{fig2}(\emph{a}) for $Re = 40$ and in Figure~\ref{fig5}(\emph{a}) for $Re = 60$ are provided as initial guesses to the Newton-GMRES hook step algorithm \citep{viswanath2007recurrent, chandler2013invariant}.  We see a convergence of the residual to $\mathcal{O}(10^{-14})$ for the $Re = 40$ case within 30 iterations as shown in Figure~\ref{fig6}(\emph{a}). The converged true equilibrium solution of the Kolmogorov flow at $Re = 40$ is shown in Figure~\ref{fig6}(\emph{b}). We can see that this solution is the same as that obtained in Figure~\ref{fig3}(\emph{d}). However, for the $Re = 60$ case, the converged true equilibrium solution obtained from the Newton search with residual of $\mathcal{O}(10^{-11})$, shown in Figure~\ref{fig6}(\emph{c}), is different from that in Figure~\ref{fig5}(\emph{b}) with the elongated filaments being absent. These patterns can also be justified from the energy spectra comparison at $Re = 40 \textup{ and } 60$ in Figures \ref{fig6}(\emph{d}) and \ref{fig6}(\emph{e}), respectively. At $Re = 40$, the energy spectra of the Kolmogorov flow solution obtained after removing the forcing term and solution from the NHK root search method are almost the same except for very tiny and negligible differences at small wave numbers. However, at $Re = 60$, the two solutions differ more due to faster energy transfer to the lower wave numbers. The differences in energy and dissipation rate values of the solutions are highlighted in Figure~\ref{fig6}(\emph{f}). We note that similar solutions for the Newton search are obtained by using the settled solution before growing perturbations take effect as the initial condition of the NKH algorithm.

\section{Conclusions and future directions}
\label{sec4}

This paper introduces a selective modification strategy aimed at steering turbulent flows toward stable flow states and identifying non-trivial steady states. This strategy autonomously pinpoints suitable forcings to alter a flow's energy and enstrophy according to a user-defined rate of modification. When applied to Kolmogorov flow, this control strategy reveals the forcing fields capable of influencing the dynamics of this unsteady turbulent flow, ultimately achieving non-trivial steady states in the modified Kolmogorov flow.

The steady-state solutions obtained are dictated by the underlying dynamics of the flow. In many cases, the dynamical system settles to a non-trivial steady state (with the laminar flow being a trivial steady state) when stabilized with the selective forcing term. This particularly happens when we selectively decrease the energy of the flow while keeping the enstrophy nominally conserved. Linear stability analysis of these solutions has unveiled that some steady-state solutions of the modified Kolmogorov flow closely approximate the invariant solutions of the actual Kolmogorov flow. By initializing the Kolmogorov flow simulations with the modified steady-states, we were able to drive the simulations toward these invariant solutions, until instabilities pushed the flow towards unsteadiness.

The demonstrated efficacy of the selective modification strategy in governing turbulent flows, particularly those exhibiting pronounced inverse energy cascades, offers a compelling case for its application in future research endeavors. The potential of this strategy to address more sophisticated turbulent flows, including those constrained by wall boundaries \cite{graham2021exact}, further underscores its significance. The introduction of wall boundaries, in contrast to the Kolmogorov flow's stationary and spatially periodic sinusoidal forcing, brings about nonlinear spatiotemporal complexities in the flow dynamics. These complexities will likely necessitate refinements in the methodology to reliably discover steady-state solutions within such constrained environments.

\begin{acknowledgments}
The authors thank Dr. Mohammad Farazmand for the feedback on the results of this manuscript.
\end{acknowledgments}

\section*{Data Availability Statement}
The data/code will be openly available on github.com/nairaditya on publication.

%\nocite{*}
\bibliography{aipsamp}% Produces the bibliography via BibTeX.

\end{document}